# Impact of Web 2.0 Technologies on Academic Libraries: A Survey on Affiliated Colleges of Solapur University

Patel Adam Burhansab
*Annamalai University*, pateladam.lib@gmail.com

M Sadik Batcha
*Annamalai University*, msbau@rediffmail.com

Muneer Ahmad
*Annamalai University*, muneerbangroo@gmail.com

# Impact of Web 2.0 Technologies on Academic Libraries: A Survey on Affiliated Colleges of Solapur University

Patel Adam Burhansab[1] Dr. M Sadik Batcha[2] Muneer Ahmad[3]

[1]*Research Scholar, Department of Library and Information Science, Annamalai University, Annamalai nagar, pateladam.lib@gmail.com*

[2]*Research Supervisor and Mentor, Professor and Librarian, Annamalai University, Annamalai nagar, msbau@rediffmail.com*

[3]*Research Scholar, Department of Library and Information Science, Annamalai University, Annamalai nagar, muneerbangroo@gmail.com*

**Abstract**

The paper aims to present the results of a survey of academic libraries about the adoption and perceived impact of Web 2.0 technologies. A total of 26 college libraries affiliated with Solapur University participated among the members. It was found that each library was using some form of technology, such as RSS, blogs, social networking sites, wikis, and instant messaging. Analyzing the entire college web technology usages, it is observed from the results that most of the web technologies are not used by the mainstream of the users due to lack of awareness, training, etc. Accurate and appropriate training should be conducted by the colleges' Libraries according to the necessities of the users. Systematic training will inevitably help the user for the maximum utilization of e-resources of the library. The leading web technologies such as internet surfing, emails, search engines, wikis, photo sharing, etc. are used by a great number of users on a daily and weekly basis of frequency. On the other hand majority of the web technologies are never used by a great number of users.

**Keywords:** Academic libraries, Websites, Wikis, Internet, ICT.

## 1. Introduction

Information Communication Technology (ICT) is one of the important buzzwords of today's IT world. The rapidgrowth of information and communication technologies have gave rise to the evolution of several new jargons like paperless society, electronic resources, portal / gateway and global digital library. In the day context, alltypes of libraries are not only providing printed resources to their library users rather they provide printed, electronic as well as other Internet resources like e-books and databases for fulfilling the day to day academic andresearch requirements of the library users (Sinha, 2012)

The Internet has provided universal access toinformation. Technological innovation has dramatically increased the rate of conversion of knowledge, information and data into electronic format. Developments in the software arena have generated powerful knowledge management software which has transformed the way knowledge is organized, stored,accessed and retrieved (Sharma, 2011).

The Internet is one of the most important and complex innovations of mankind. It is a powerful means of communication, dissemination and retrieval of information. It is a network of network connecting thousands of smaller computer networks together so that other networks may share information present in one network. It is one of the powerful / effective tools or technologies ever produced for getting information on fingertips from any part of the world even sitting at one's own location. Internet has been described as a system for allowing computers to communicate with each other. It is used by millions of people throughout the world for communication, business, research, recreation and browse information for higher studies. There are various national, international/global networks systems, morethan 40,000, readily accessible through the Internet. As network bandwidths increases it will become common to have video and animation over networks, there by challenging the conventional analog media such as cable TV and videotapes. Now the facility of Internet has been increasingly used for educational course delivery (Sinha, 2004).

The Internet facility in India has grown tremendously over the years. The efficiency and capabilities in providing right information to right person at right time has grown rapidly with the use of Internet. The recent phenomenon and the emergence of information society, knowledge driven economy and the cost effectiveness of technological gadgets has made Internet an unavoidable necessity for every institutions of higher learning and research (Sharma, Singh and Mishra, 2008). The Internet tools and services have been used by the teachers' scholars and students for variety of purposes for day to day academic and research activities.

## 2. Review of Literature

Miranda et al. (2010) mentioned the pros of Web 2.0 for librarians. They include "collaboration, customization, communication, knowledge generation, sharing, updating, flexible tools, speed, reduction of costs, training, and facilitates experimentation" They also mentioned pros of such technologies for library users.They are "low level of complexity, requires little technical expertise, reduction of costs, flexibility, user involvement, time saving, reduces information overload, social tagging, idea sharing, and knowledge generation and sharing". In a study of institutional repository at Columbia University, (Cocciolo, 2010) established that the application of Web 2.0 in designing the project garnered significantly more community participation as user contributions increased by 9,728 percent. (O'Dell, 2010) mentioned that with the use of new social media, library staff enjoyed professional development opportunities at a very low cost. (Nogueira, 2010) noted the benefit of the use of Web 2.0 technologies as the "growing number of publicthey reach (visitors, potential users or actual users)". Issues in implementing Web 2.0 technologies in Chinese libraries, mentioned by (Cao, 2009) include management buy-in, lack of awareness, lack of user participation, andlack of technology staff. (Garcia-Perez and Ayres, 2010) surveyed the users of a wiki about the reasons of decline in its use. The answers included lack of time, lack of interest/need, and lack of direct accessibility through a desktop or homepage. In a survey by (Chawner, 2008), LIS professionals in New Zealand mentioned problems like doubtful quality of information, privacy and security issues, increasing rate of change and challenge of keeping up, and lack of staff training. Based on a survey of Web 2.0 implementation in information literacy instruction, (Luo, 2010) stated the problems to be a lack of skilled staff, a lack of Web 2.0 knowledge among students, and online vandalism in wikis. In a focus group study conducted by (Burhanna et al. 2009), students mentioned problems of authority of information and the privacy of users. (Burhansab, Batcha & Ahmad, 2020) investigated the use of electronic resources/information by library users in selected colleges of Solapur University. Specifically, to investigate the awareness and level of use of electronic resources; perceived reliance, benefits and impact of use of electronic resources on the research activities. The research design for the study was a survey. Questionnaire schedule was used to collect data from 1022 library users from selected colleges of Solapur University. The result revealed that preponderance of users from aided 33.51% Self-financing 26.10% and Education colleges 43.24 % preferred to visit the Library once in three days. While analyzing the entire college libraries regarding the frequency of visit, users gave first preference to once in three days i.e. 27.2%. College wise analysis reveals that mainstream of users from Aided Colleges 38%, Self-financing Colleges 28.3%, Engineering Colleges 43%, Education colleges 53.2% and Pharmacy Colleges 23.4% are spending their time 1-2 hrs. in libraries and 40.8% visit college libraries to issue and return books and in the device usage (33.9%) of users ranked mobile phone as the second device for accessing the e-resources. It is noticeable that 24.7% of users acknowledge from Aided Colleges know about the services of the Library from website but in Self Finance Colleges 8.5% and Pharmacy College 56.4% users aware about the services from the friends it is observed that most of the web technologies are not used by the mainstream of the users due to lack of awareness, training etc.

## 3. Objectives

There are various technologies that have been grouped as Web 2.0, the second version of the worldwide web. As in other spheres of life, these technologies have their place inacademic libraries. Libraries and their users are enjoying the benefits of Web 2.0 technologies but they are also prone to problems and threats. However, no empirical data are available on the impact of such technologies in libraries. In order to improve the useof these technologies in libraries, there is a need to study users' perceptions (whether positive or negative) of Web 2.0 technologies with regard to academic library services. This study was conducted to accomplish the following objectives: to identify the adoption of various Web 2.0 technologies in academic libraries; and to articulate the perceptions of library users about advantages and disadvantages of these technologies for libraries.

## 4. Research Methodology

### 4.1. Sampling and Questionnaire Design

A survey was conducted in 26 colleges of Solapur University based in Maharashtra, India. A structured questionnaire was designed. Library users within each institute were randomly. A total of one thousand and twenty two (1022) library users were randomly selected from the 26 institutes. The questionnaire consisted of various questions in the following categories: Demographics, use of e-resources, access to e-resources, competencies and training, challenges and benefits of e-resources among others. There were also two open-ended question that asked respondents about the impediment and improvement to e-resources in their institutes' libraries. In all 1050 questionnaires were distributed to the library users but 1022 were received. The response rate was 97 per cent. Statistical package for social sciences (SPSS version 16) was used to analyse the data.

### 4.2. Study Population

Out of 114 affiliated colleges to Solapur University, 26 colleges (including Aided Self, Financing Colleges, Engineering Colleges, Education Colleges & Pharmacy Colleges are selected for the study. The study population constituted library users of 26 college Libraries affiliated with Solapur University, Solapur.

### 4.3. Sample Selection

Sample is selected on the basis of confidence level 99% and confidence interval 5%. Following are the samples of the study of various colleges affiliated to Solapur University, Solapur. The sample and total population taken for the study are shown in Tabulation (Table 4.3)

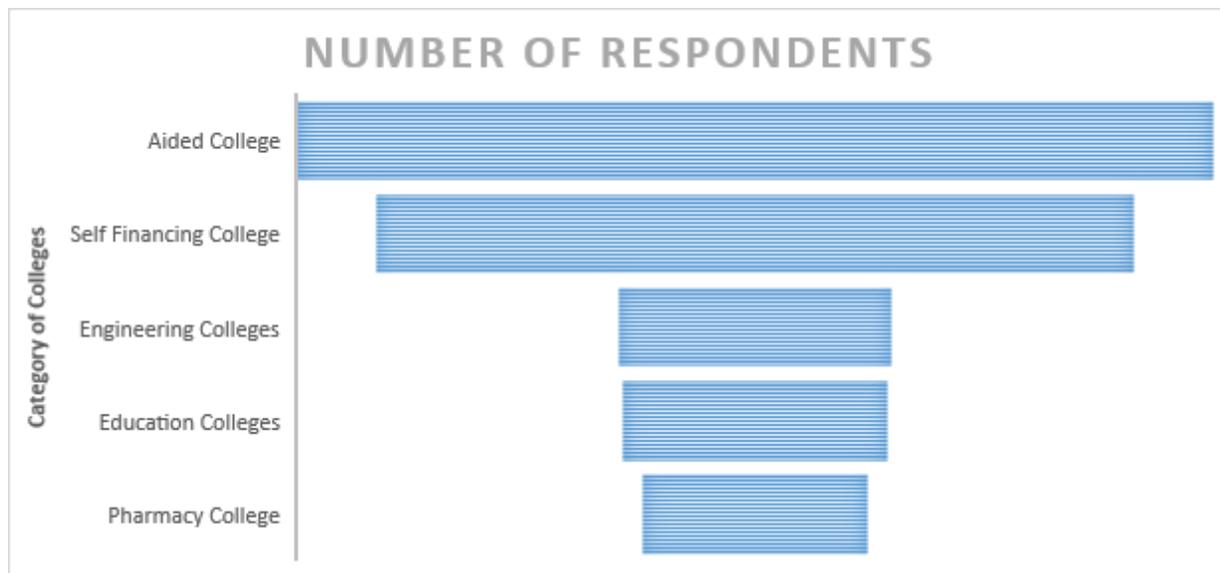

**Table 4.3: Representation of Respondents from Different Categories of colleges**

| S.No | Category of the Colleges | Male | Female | Total |
|---|---|---|---|---|
| 1. | Aided College | 245 | 140 | 385 |
| 2. | Self Financing College | 174 | 144 | 318 |
| 3. | Engineering Colleges | 67 | 47 | 114 |
| 4. | Education Colleges | 66 | 45 | 111 |
| 5. | Pharmacy College | 59 | 35 | 94 |
|  | **Total** | **611** | **411** | **1022** |

## 5. Results and Discussions

### 5.0. College-wise analysis of frequency of using various web technologies in is enlightened

Analyzing the entire college web technology usages, it is observed from the tables below that most of the web technologies are not used by the mainstream of the users due to lack of awareness, training etc. Accurate and appropriate training should be conducted by the colleges Libraries according to the necessities of the users. Systematic training will inevitably help the user for maximum utilization of e-resources of the library. The leading web technologies such as internet surfing, emails, search engines, wikis, photo sharing etc. are used by the great number of users on daily and weekly basis of frequency. On the other hand majority of the web technologies are never used by the great number of users.

**Table 5.1: Internet surfing**

| Colleges / Frequency | Aided Colleges | | Self Finance College | | Engineering Colleges | | Education Colleges | | Pharmacy Colleges | | Total | |
|---|---|---|---|---|---|---|---|---|---|---|---|---|
| Daily | 152 | 39.5 % | 94 | 29.6 % | 22 | 19.3 % | 48 | 43.2 % | 64 | 68.1 % | 380 | 37.2 % |
| Weekly | 90 | 23.4 % | 99 | 31.1 % | 39 | 34.2 % | 25 | 22.5 % | 7 | 7.4 % | 260 | 25.4 % |
| Monthly | 40 | 10.4 % | 40 | 12.6 % | 12 | 10.5 % | 8 | 7.2 % | 6 | 6.4 % | 106 | 10.4 % |
| Occasionally | 31 | 8.1 % | 31 | 9.7 % | 11 | 9.6 % | 12 | 10.8 % | 9 | 9.6 % | 94 | 9.2 % |
| Never | 72 | 18.7 % | 54 | 17.0 % | 30 | 26.3 % | 18 | 16.2 % | 8 | 8.5 % | 182 | 17.8 % |
| Total | 385 | 100 % | 318 | 100 % | 114 | 100 % | 111 | 100 % | 94 | 100 % | 1022 | 100 % |

Chi Square = 279.9  P Value = 0.000

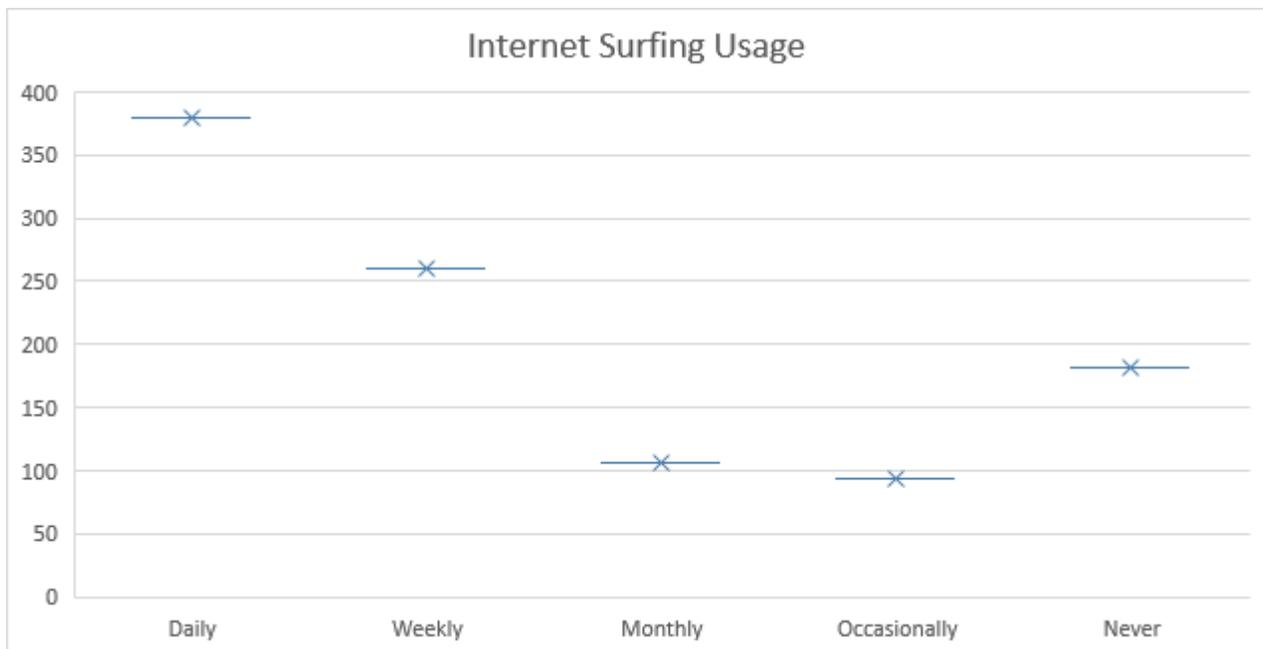

Majority of the users from the Aided Colleges (39.5%), Self Financing Colleges (29.6%), Education Colleges (43.2%), and Pharmacy Colleges (68.1%) are using Internet facilities on daily basis. But greater part of the users from Engineering Colleges (19.4%) is familiar with surfing Internet on weekly basis. Yet the internet facilities are found never used by a group which is higher in number i.e. Aided Colleges (18.7%), Self Financing College (17.0%), Engineering Colleges (26.3%), Education Colleges (16.2%), and Pharmacy College (8.5%). Chi-square value 279.9 and p-value zero indicate that there is a significant difference in the frequency of internet suffering among the users in various libraries affiliated to Solapur University.

Scrutinizing the entire college, the internet surfing on daily basis (37.2%) are found more followed by weekly basis (25.4%). Apart from this, 17.8% of users have said they never used this facility where as the users those use found on monthly wise (10.4%) and occasionally (9.2%) which are lesser in number.

**Table 5.2: Search Engines**

| Colleges / Frequency | Aided Colleges | | Self Finance College | | Engineering Colleges | | Education Colleges | | Pharmacy Colleges | | Total | |
|---|---|---|---|---|---|---|---|---|---|---|---|---|
| Daily | 110 | 28.6 % | 115 | 36.2 % | 38 | 33.3 % | 51 | 45.9 % | 62 | 66.0 % | 376 | 36.8 % |
| Weekly | 159 | 41.3 % | 102 | 32.1 % | 30 | 26.3 % | 22 | 19.8 % | 9 | 9.6 % | 322 | 31.5 % |
| Monthly | 27 | 7.0 % | 33 | 10.4 % | 20 | 17.5 % | 8 | 7.2 % | 7 | 7.4 % | 95 | 9.3 % |
| Occasionally | 28 | 7.3 % | 28 | 8.8 % | 14 | 12.3 % | 12 | 10.8 % | 10 | 10.6 % | 92 | 9.0 % |
| Never | 61 | 15.8 % | 40 | 12.6 % | 12 | 10.5 % | 18 | 16.2 % | 6 | 6.4 % | 137 | 13.4 % |
| Total | 385 | 100 % | 318 | 100 % | 114 | 100 % | 111 | 100 % | 94 | 100 % | 1022 | 100 % |

Chi Square = 279.9   P Value = 0.000

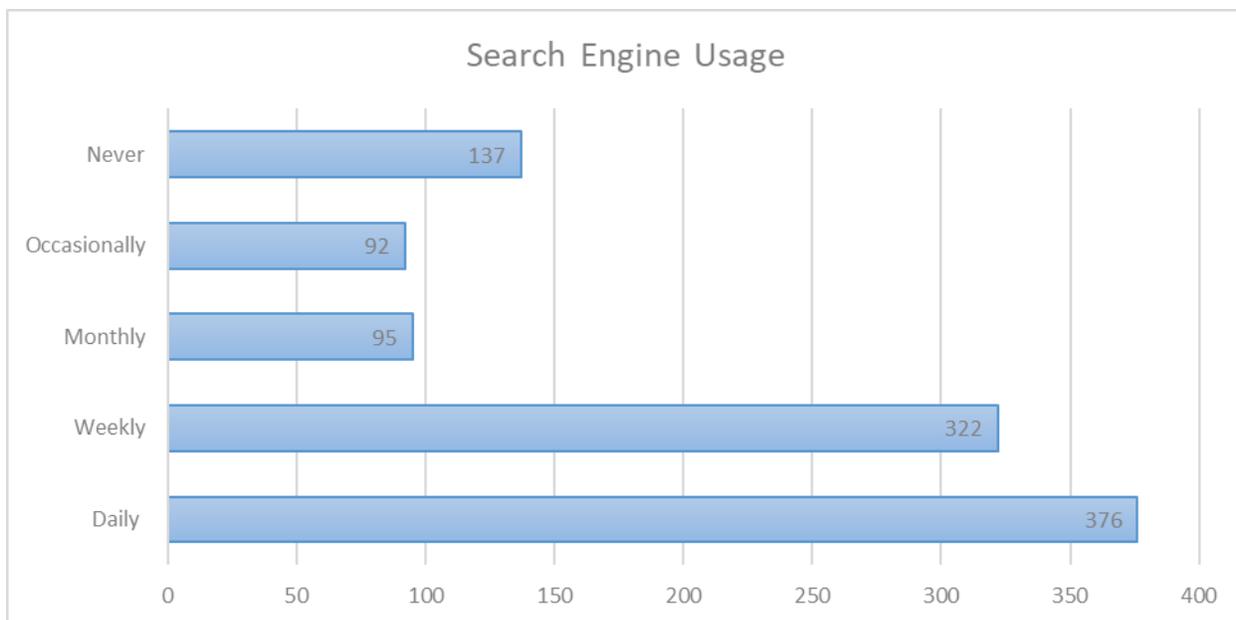

Like Internet surfing the search engines are also mainly used by users on daily basis in self financing colleges (36.2%), Engineering Colleges (33.3%), Education Colleges (45.9%), and Pharmacy College (66.0%). The same is used by other group of users on weekly basis in Aided Colleges (42.1%). Still higher number of users are noted in never use of search Engines. They are from Aided Colleges (98.0%), Self Financing College (12.6%), Engineering Colleges (10.5%), Education Colleges (16.3%), and Pharmacy College (6.4%). The Chi-square value 235.5 and the p-value zero indicate that there is a significant different in the frequency of using search engines among the users in various colleges libraries affiliated to Solapur University.

Considering the search Engine usage of whole colleges, the outcomes make to known that it may be used more on daily (37.1%) and followed by weekly (31.8%). However, a considerable number of users never (11.2%) used these facilities. Targeting this group necessary measure may be taken by the authorities to improve the usage.

Table 5.3: Web Technology Usage - Emails

| Colleges / Frequency | Aided Colleges | | Self Finance College | | Engineering Colleges | | Education Colleges | | Pharmacy Colleges | | Total | |
|---|---|---|---|---|---|---|---|---|---|---|---|---|
| Daily | 142 | 36.9 % | 138 | 43.4 % | 38 | 33.3 % | 56 | 50.5 % | 59 | 62.8 % | 433 | 42.4 % |
| Weekly | 102 | 26.5 % | 63 | 19.8 % | 31 | 27.2 % | 22 | 19.8 % | 12 | 12.8 % | 230 | 22.5 % |
| Monthly | 57 | 14.8 % | 51 | 16.0 % | 19 | 16.7 % | 8 | 7.2 % | 6 | 6.4 % | 141 | 13.8 % |
| Occasionally | 35 | 9.1 % | 42 | 13.2 % | 19 | 16.7 % | 14 | 12.6 % | 10 | 10.6 % | 120 | 11.7 % |
| Never | 49 | 12.7 % | 24 | 7.5 % | 7 | 6.1 % | 11 | 9.9 % | 7 | 7.4 % | 98 | 9.6 % |
| Total | 385 | 100 % | 318 | 100 % | 114 | 100 % | 111 | 100 % | 94 | 100 % | 1022 | 100 % |
| Chi Square = 134.0 | | | | | | | | | | | P Value = 0.000 | |

It is observed that E-mails are mainly checked by the greater part of users on daily basis in Aided Colleges (36.9%), Self Financing colleges (43.4%), Engineering colleges (33.3%), Education Colleges (50.5%), and Pharmacy College (62.8%). The weekly usage of email facility by the users of Aided College (26.5%), Self-Financing Colleges (19.8%), Engineering Colleges (27.2%), Education College (19.8%), and Pharmacy College (12.8%) are at the next level. Chi-square value 134.0 and the p-value zero indicate that there is a significant difference in the frequency of using emails among the users in various colleges affiliated to Solapur University. Overall analysis shows that 42.4% users on daily basis and 22.5% users on weekly basis check their emails. Only a smaller group alone found never (9.6%) using e-mails. Apart from this, monthly wise (13.8%) and occasionally (11.7%) are also noted in higher number.

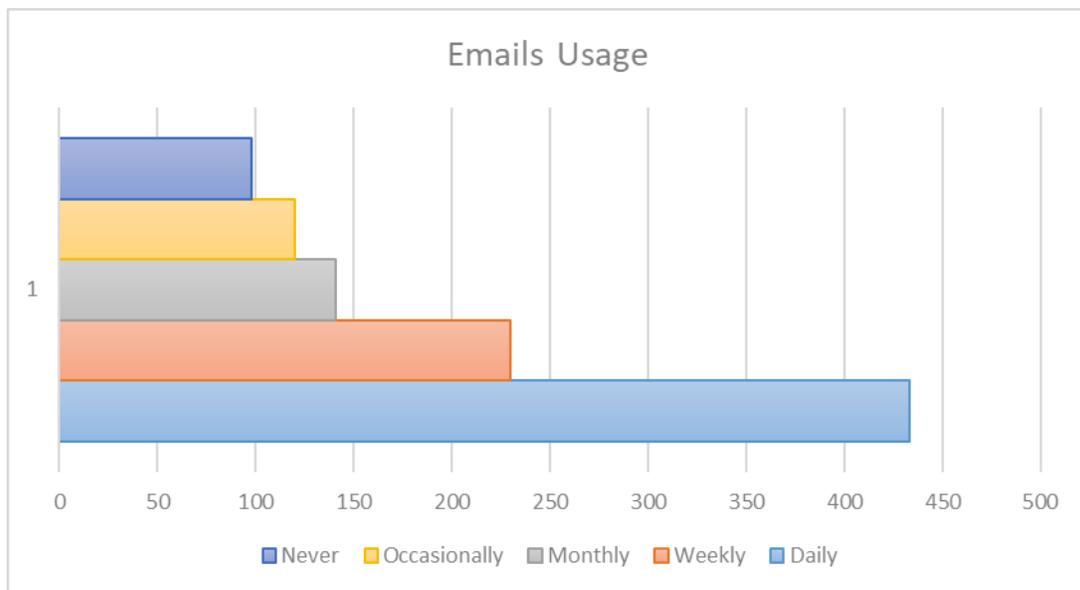

**Table 5.4: Blogs-Web Technology Usage**

| Colleges / Frequency | Aided Colleges | | Self Finance College | | Engineering Colleges | | Education Colleges | | Pharmacy Colleges | | Total | |
|---|---|---|---|---|---|---|---|---|---|---|---|---|
| Daily | 44 | 11.4 % | 28 | 8.8 % | 15 | 8.8 % | 22 | 17.2 % | 25 | 26.6 % | 134 | 12.9 % |
| Weekly | 70 | 18.2 % | 62 | 19.5 % | 19 | 16.7 % | 26 | 20.3 % | 27 | 28.7% | 204 | 19.6 % |
| Monthly | 103 | 26.8 % | 50 | 15.7 % | 15 | 13.2 % | 15 | 11.7 % | 10 | 10.6% | 193 | 18.6 % |
| Occasionally | 26 | 6.8 % | 33 | 10.7 % | 16 | 14.0 % | 16 | 12.5% | 19 | 20.2 % | 110 | 10.6 % |
| Never | 142 | 36.9 % | 145 | 45.6 % | 49 | 43.0 % | 49 | 38.3% | 13 | 13.8 % | 398 | 38.3 % |
| Total | 385 | 100 % | 318 | 100 % | 114 | 100 % | 111 | 100 % | 94 | 100 % | 1022 | 100 % |

Chi Square = 294.9                                                           P Value = 0.000

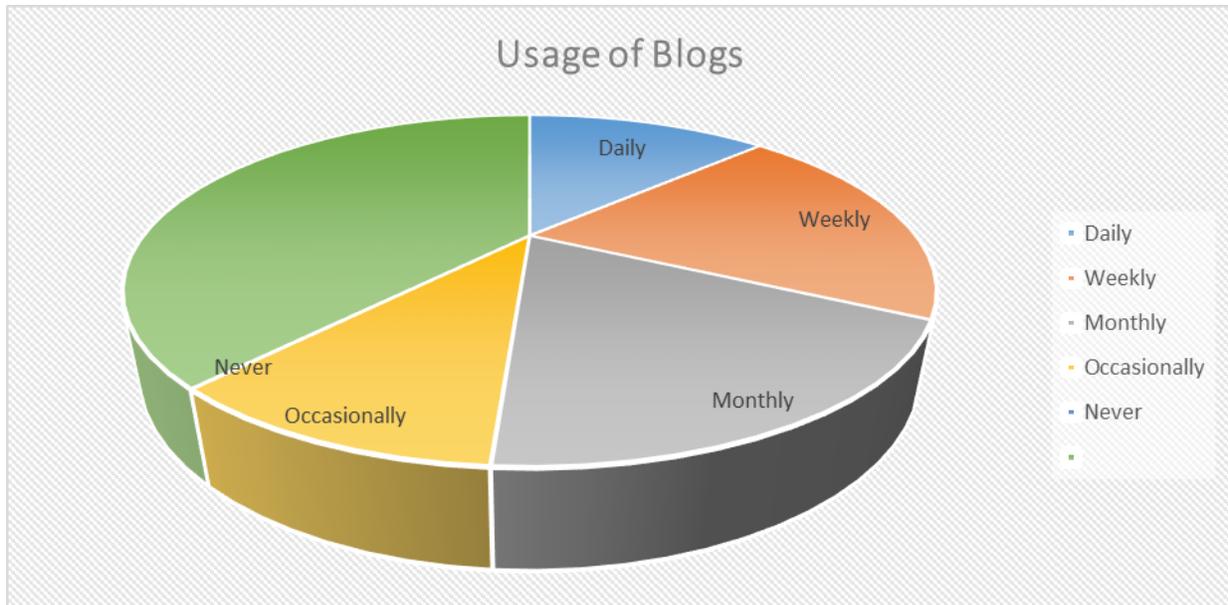

While analyzing the above table it is observed that the majority of the users from Aided College (26.5%) use blogs on monthly basis and in Pharmacy College (20.2%) they are using occasionally. In addition to the above, the users from Self-financing College (19.5%), Engineering Collages (16.6%), and Pharmacy Colleges (28.8%) are also using blogs on weekly basis. A set of users from Self Financing colleges (45.5%), and Engineering Colleges (42.9%) are noted never using blogs which is higher in number. Chi-square 294.9 and the p-value zero indicate that there is a significant difference in the frequency of using blogs among the users in various libraries affiliated to Solapur University. Overall analysis reveals that 36.3% never used blogs. However positively 19.8% users used blogs on weekly and another 18.8% of users used it on monthly-wise frequency.

**Table 5.5: E-Resource Search**

| Colleges / Frequency | Aided Colleges | | Self Finance College | | Engineering Colleges | | Education Colleges | | Pharmacy Colleges | | Total | |
|---|---|---|---|---|---|---|---|---|---|---|---|---|
| Daily | 42 | 10.9 % | 20 | 6.3 % | 19 | 16.7 % | 21 | 18.9 % | 28 | 29.8 % | 130 | 12.7 % |
| Weekly | 105 | 27.3 % | 87 | 27.4 % | 15 | 13.2 % | 41 | 36.9 % | 28 | 29.8% | 276 | 27.0 % |
| Monthly | 62 | 16.1 % | 87 | 27.4 % | 19 | 16.7 % | 14 | 12.6 % | 10 | 10.6 % | 192 | 18.8 % |
| Occasionally | 37 | 9.6% | 44 | 13.8 % | 19 | 16.7 % | 22 | 19.8 % | 18 | 19.1 % | 140 | 13.7 % |
| Never | 139 | 36.1 % | 80 | 25.2 % | 42 | 36.8 % | 13 | 11.7 % | 10 | 10.6 % | 284 | 27.8 % |
| Total | 385 | 100 % | 318 | 100 % | 114 | 100 % | 111 | 100 % | 94 | 100 % | 1022 | 100 % |

Chi Square = 307.8                                                                                          P Value = 0.000

The above table 5.5 explains Searching E-resources by the users of various libraries. It indicates that greater part of the users from the Aided College (36.1%), and Engineering College (36.8%) are observed never go for searching E-resources but at the same time weekly usage is found among the users from Aided Colleges (27.3%), from Engineering College (13.2%). They are calculated at considerable in number. However, in the case of Self-financing College 27.4% students use it on monthly basis and 27.4% students on weekly basis. In Education Collages (36.9%) the students use it on weekly and in Pharmacy Colleges (29.8%) the students make use it on weekly and (29.8%) also on daily.

Overall analysis of the whole libraries explains that majority of users are never (27.8%) going for search of E-resources. But at the same time 27.0% users use it on weekly and 18.8% users on monthly basis. Chi-square 307.8 and the p-value zero indicate that there is a significant difference in the frequency of using E-resources search among the users in various libraries affiliated to Solapur University.

Table 5.6: Downloading of E-documents

| Colleges / Frequency | Aided Colleges | | Self Finance College | | Engineering Colleges | | Education Colleges | | Pharmacy Colleges | | Total | |
|---|---|---|---|---|---|---|---|---|---|---|---|---|
| Daily | 65 | 16.9 % | 34 | 10.7 % | 15 | 13.2 % | 33 | 29.7 % | 34 | 36.2 % | 181 | 17.7 % |
| Weekly | 89 | 23.1 % | 81 | 25.5 % | 34 | 29.8 % | 26 | 23.4 % | 21 | 22.3% | 251 | 24.6 % |
| Monthly | 66 | 17.1 % | 26 | 8.2 % | 15 | 13.2 % | 11 | 9.9 % | 6 | 6.4 % | 124 | 12.1 % |
| Occasionally | 37 | 9.6 % | 65 | 20.4 % | 16 | 14.0 % | 15 | 13.5 % | 27 | 28.7 % | 160 | 15.7 % |
| Never | 128 | 33.2 % | 112 | 35.2 % | 34 | 29.8 % | 26 | 23.4 % | 6 | 6.4 % | 306 | 29.9 % |
| Total | 385 | 100 % | 318 | 100 % | 114 | 100 % | 111 | 100 % | 94 | 100 % | 1022 | 100 % |

Chi Square = 307.8                                                                                          P Value = 0.000

The students who are never downloading the E-resources are found high in Aided Colleges (33.2%), Self Financing College (35.2%), Engineering College (29.8%), Education College (23.4) and Pharmacy College (6.4%). However, 17.1% students from Aided College use this facility on monthly basis. Remarkable number of students from Aided colleges (23.1%), Self Financing College (25.5%), Engineering College (29.8%), Education Colleges (23.4%), and Pharmacy College (22.3%) use it on weekly basis. In addition, students from Aided Colleges

(9.6%), and Self-Financing College (20.4%) are found occasionally using the e-resources downloading facilities. Chi-square value 309.4 and p-value zero indicate that there is a significant difference in the frequency of downloading E-documents among the students in various libraries affiliated to Solapur University.

Overall analysis of the entire libraries explains that mainstream of the students are observed never (29.9%) using the download of E-resources. Yet at the same time 24.6% users use it on weekly and 17.7% are on daily basis.

**Table 5.7: RSS feeds**

| Colleges / Frequency | Aided Colleges | | Self Finance College | | Engineering Colleges | | Education Colleges | | Pharmacy Colleges | | Total | |
|---|---|---|---|---|---|---|---|---|---|---|---|---|
| Daily | 33 | 8.6 % | 29 | 9.1 % | 19 | 16.7 % | 18 | 16.2 % | 15 | 16.0 % | 114 | 17.7 % |
| Weekly | 69 | 17.9 % | 43 | 13.5 % | 19 | 16.7 % | 22 | 19.8 % | 32 | 34.0 % | 185 | 24.6 % |
| Monthly | 25 | 6.5 % | 36 | 11.3 % | 8 | 7.0 % | 33 | 29.7 % | 11 | 11.7 % | 113 | 12.1 % |
| Occasionally | 23 | 6.0 % | 22 | 6.9 % | 23 | 20.2 % | 12 | 10.8 % | 14 | 14.9 % | 94 | 15.7 % |
| Never | 235 | 61.0 % | 188 | 59.1 % | 45 | 39.5 % | 26 | 23.4 % | 22 | 23.4 % | 516 | 29.9 % |
| Total | 385 | 100 % | 318 | 100 % | 114 | 100 % | 111 | 100 % | 94 | 100 % | 1022 | 100 % |

Chi Square = 334.6      P Value = 0.000

Rich site summary, originally RDF Site summery often called really simple syndication is never used by the major part of the students from Aided Colleges (61.0%), Self Financing College (59.1%), and Engineering Colleges (39.4%). However the (29.7%) the students from Education Colleges preferred it on monthly scale and another set of users choose (19.8%) it weekly scale. In Pharmacy College 34.0% students choose RSS feed at weekly frequency and also 16.0% say with daily usage. Chi-square value 334.6 and p-value zero indicate that there is a significant difference in the frequency of using RSS feeds among the students in various libraries affiliated to Solapur University.

Overall analysis of the all colleges illustrate that 50.5% of users never use the facilities but apart from this 18.1% of users utilize it on weekly and 11.2% users use it on daily basis.

**Table 5.8: Mobile / E-mail alert Services**

| Colleges / Frequency | Aided Colleges | | Self Finance College | | Engineering Colleges | | Education Colleges | | Pharmacy Colleges | | Total | |
|---|---|---|---|---|---|---|---|---|---|---|---|---|
| Daily | 56 | 14.5 % | 30 | 9.4 % | 19 | 16.7 % | 22 | 19.8 % | 30 | 31.9 % | 157 | 15.4 % |
| Weekly | 116 | 30.1 % | 68 | 21.4 % | 34 | 29.8 % | 15 | 13.5 % | 15 | 16.0 % | 248 | 24.3 % |
| Monthly | 62 | 16.1 % | 34 | 10.7 % | 16 | 14.0 % | 14 | 12.6 % | 10 | 10.6 % | 136 | 13.3 % |
| Occasionally | 26 | 6.8 % | 34 | 10.7 % | 14 | 12.3 % | 15 | 13.5 % | 14 | 14.9 % | 103 | 10.1 % |
| Never | 125 | 32.5 % | 152 | 47.8 % | 31 | 27.2 % | 45 | 40.5 % | 25 | 26.6 % | 378 | 37.0 % |
| Total | 385 | 100 % | 318 | 100 % | 114 | 100 % | 111 | 100 % | 94 | 100 % | 1022 | 100 % |

Chi Square = 224.6     P Value = 0.000

Table 5.8 elucidates that majority of the users from Aided College (32.5%) and Self-Financing College (47.8%) never gets use of mobile /E-mail alert services. But in Aided Colleges (29.9%), Self Financing College (21.4%) and Engineering Colleges (29.8%) users make use the benefit of mobile/ e-mail alert services which are found on weekly basis. In case of Education Colleges 40.5% users never use the benefits of Mobile/ e-mail alert services.

Preponderance of the users selected 'never' category in case of from Self Financing colleges (47.8%), and Education colleges (40.5%). On the other hand a remarkable number of users from Self Financing Colleges (21.4%) and Engineering Colleges (29.8%) get use of mobile/ e-mail alert services on weekly basis and in Pharmacy College 32% users get it on daily basis. Chi-square value 224.6 and the p value zero indicate that there is a significant difference in the frequency of using Mobile alert/ E-mail alert services among the users in various libraries affiliated to Solapur University.

Overall analysis of the all colleges demonstrates that 37.0% never use the facilities, apart from this 24.3% of users get it on weekly and 15.4% users get it at daily basis.

**Table 5.9: Wiki**

| Colleges / Frequency | Aided Colleges | | Self Finance College | | Engineering Colleges | | Education Colleges | | Pharmacy Colleges | | Total | |
|---|---|---|---|---|---|---|---|---|---|---|---|---|
| Daily | 53 | 13.8 % | 64 | 20.1 % | 22 | 19.3 % | 15 | 13.5 % | 40 | 42.6 % | 194 | 19.0 % |
| Weekly | 122 | 31.7 % | 89 | 28.0 % | 38 | 33.3 % | 40 | 36.0 % | 23 | 24.5 % | 312 | 10.5 % |
| Monthly | 68 | 17.7 % | 27 | 8.5 % | 8 | 7.0 % | 16 | 14.4 % | 9 | 9.6 % | 128 | 12.5 % |
| Occasionally | 31 | 8.1 % | 30 | 9.4 % | 19 | 16.7 % | 18 | 16.2 % | 12 | 12.8 % | 110 | 10.8 % |
| Never | 111 | 28.8 % | 108 | 34.0 % | 27 | 23.7 % | 22 | 19.8 % | 10 | 10.6 % | 278 | 27.2 % |
| Total | 385 | 100 % | 318 | 100 % | 114 | 100 % | 111 | 100 % | 94 | 100 % | 1022 | 100 % |

Chi Square = 237.5                                                                                           P Value = 0.000

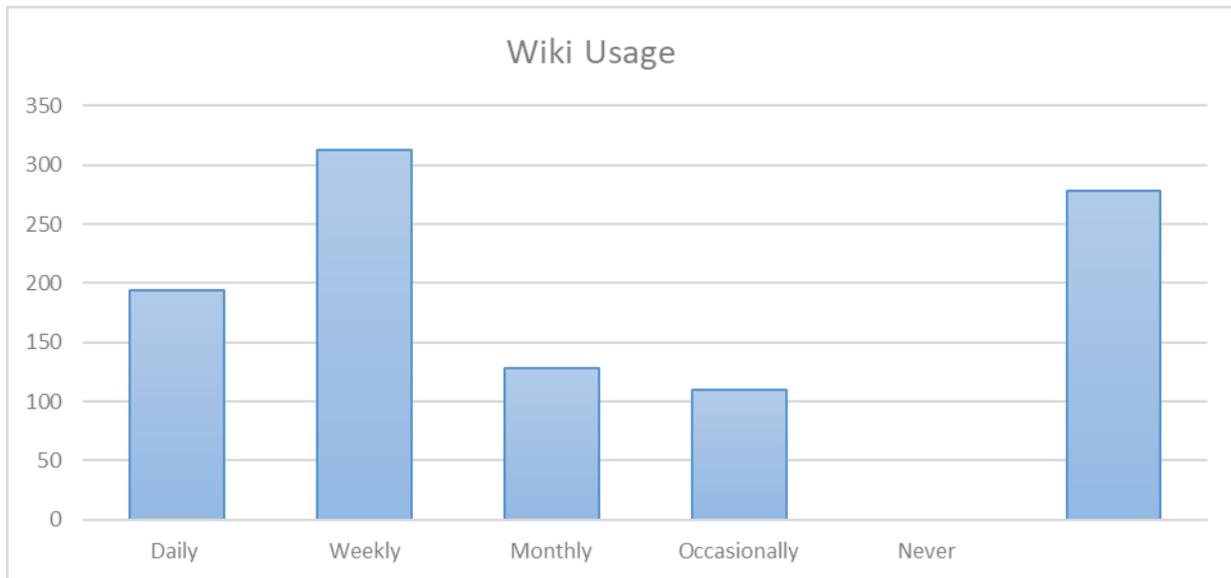

It is obvious from the table 5.9 that best part of the users from Aided Colleges (31.7%), engineering colleges (33.3%) and Education Colleges (36.0%) support Wiki at weekly usage. But the preponderance of users from Pharmacy Colleges (42.5%) selects it at daily option. In self financing collages (34%) majority of users are not using wiki. At the same time more number of daily users from self financing colleges (20.1%) using Wiki. Weekly users from Education colleges (36.0%) and Engineering Colleges (33.3%) are also higher in numbers. Chi-square value 237.5 and p-value zero indicate that there is a significant difference in the frequency of using Wiki among the users in various libraries affiliated to Solapur University.

Overall analysis reveals that wiki the content management system is used by the majority of the users on weekly (30.5%) frequency and about 19.0% of users use it on daily basis. But 27.2% users never use wiki.

## 6. Conclusion

The results of this study reveal that all participant academic libraries were using someform of Web 2.0 technologies. However, there was a difference in their adoption. Sometools were more popular than others. That library user favors the advantages of some toolsmore strongly reveals the potential these technologies have shown in improving libraryservices. Library users who have experienced more forms of such technologies showed more agreement with their positive features. Although the majority of the library users' didnot agree with most of negative features, there were some problems that gained theattention of a reasonable number of participants. These include doubts regarding thelongevity of tools, lack of staff time for maintenance, lack of standardization, institutional barriers in implementation, lack of staff training, and information overload.There is a need to promote the potential these technologies for libraries. The preferencesof library users that surfaced in this study can be used for this purpose. Libraries worldwidecan follow the best practice models in using Web 2.0 technologies. It is also necessarythat decision makers and technologists overcome the problems in their properimplementation and the threats that these technologies pose to their users.